\documentclass[aps,twocolumn,showpacs]{revtex4}
\usepackage[utf8]{inputenc}
\usepackage{graphicx}
\usepackage{amsmath}
\usepackage{amsfonts}
\usepackage{amssymb,color}
\usepackage{xcolor}
\usepackage{soul,xcolor}

\begin{document}

\setstcolor{red}

\title{Characterizing a transition from limited to unlimited diffusion in energy for a time-dependent billiard}

\author{Felipe Augusto O. Silveira$^{1}$, Anne Kétri P. da Fonseca$^{1}$, Peter Schmelcher$^2$, Denis G. Ladeira$^{3}$ and  Edson D.\ Leonel$^{1}$}
\affiliation{
$^{1}$ 
Departamento de Física, UNESP -- Universidade Estadual Paulista -- Av. 24A, 1515, Bela Vista, CEP: 13506-900 -- Rio Claro -- São Paulo -- Brazil
\\
$^{2}$
Institut für Laserphysik -- Universität Hamburg --  Luruper Chaussee 149 --
22761 Hamburg
\\
$^{3}$ 
Departamento de Estatística, Física e Matemática, 
UFSJ -- Universidade Federal de São João del-Rei -- Rod. MG 443, Km 7, Fazenda do Cadete, CEP: 36420-000 -- Ouro Branco -- Minas Gerais -- Brazil
}

\date{\today}\widetext

%\pacs{05.45.-a, 05.45.Pq}

\begin{abstract}

We explore Fermi acceleration in a driven oval billiard which shows unlimited to limited
diffusion in energy when passing from the free to the dissipative case. We provide evidence
for a second-order phase transition taking place while detuning the corresponding restitution
coefficient from one responsible for the degree of dissipation. A corresponding order
parameter is suggested, and its susceptibility is shown to diverge at the transition point.
We also discuss the underlying symmetry breaking and the elementary excitation of the controlled
diffusion process.

\end{abstract}

\maketitle

\section{Introduction}
Billiards represent dynamical systems composed of particles moving inside a closed boundary while colliding with the latter \cite{billiardWorks,Chernov}. Generally, the border may assume many geometrical shapes and can even depend on time. It is well-known that a phenomenon denoted as Fermi acceleration (FA) is observed for time-dependent boundaries. In the case of FA, an ensemble of particles acquires, on average, unlimited energy growth from collisions with the border. The Loskutov-Ryabov-Akinshin (LRA) conjecture \cite{LRA1,LRA2} states that chaos in a static billiard is a sufficient condition to observe FA when a time-dependent perturbation to the boundary is introduced. 

Examples of time-dependent billiards exhibiting FA include the Lorentz gas \cite{lorentzGas1,lorentzGas2}, the stadium billiard \cite{stadium}, and the oval billiard \cite{ovalb}. A well-known example complementing the LRA conjecture is the elliptical billiard \cite{robusto1}. The static version of the elliptical billiard is integrable. When a time-perturbation to the boundary is introduced, depending on the initial condition as well as the control parameters, FA is observed. Due to the boundary's time variation, the separatrix present in the phase space is replaced by a stochastic layer. Therefore, trajectories confined inside the separatrix (librators) can explore regions outside the separatrix (rotators) and vice-versa. These crossings define a mechanism producing the unlimited energy growth \cite{robusto1,robusto2}. It has been noticed that FA is not robust under inelastic collisions \cite{robusto3} since even for the smallest amount of dissipation, FA is suppressed.
For the dissipative billiard, the appearance of attractors leads the mean velocity to enter a saturation regime for large enough times, and the energy gain of the particles becomes limited \cite{Termal}.

For many nonlinear systems, some physical observables obey properties linked to scale invariance that inevitably lead to scaling laws \cite{eins,zwei,drei,vier}, which are commonly related to phase transitions \cite{funf,sechs}. Although scale invariance and power laws have been found in many different 
nonlinear dynamical systems, little is known about possible related phase transition.
In statistical physics, phase transitions are often related to changes in the
spatial structure of a system caused by the variation of control parameters \cite{sieben,arch}. Conversely, in dynamical systems, phase transitions are linked to changes in their phase space structure, also due to changes in the corresponding control parameters \cite{neun,zehn}. Close to a phase
transition, the relevant observables obey a scaling behavior and exhibit critical
exponents that characterize the system's dynamics.

In this work, we study a transition from limited to unlimited energy growth in a time-dependent oval billiard whose boundary moves in time. The particle, or in an equivalent way, an ensemble of non-interacting particles, collides with the moving boundary. The reflection law considers inelastic collisions that preserve the tangential momentum but not the kinetic energy upon collision. We assume only the normal component of the velocity to be affected by the dissipation, which is controlled by a restitution coefficient. The static oval billiard exhibits a mixed-phase space; therefore, the LRA conjecture applies. With an increasing number of collisions, the growth of the average velocity of the particles is unbounded, leading to the unlimited diffusion of energy. Dissipation occurs when the restitution coefficient is less than one, and unbounded energy growth is no longer observed. Hence the dynamics of the average velocity evolves to a stationary state, confirming the suppression of the unlimited energy growth. This suppression of unlimited growth marks the transition we are interested in. When the dissipation parameter approaches one continuously, the average velocity saturation plateau increases indefinitely. The inverse of the saturation plateau approaches zero at this transition and, as we shall see, defines a candidate for an order parameter. At the transition point, the corresponding susceptibility diverges. Therefore, it provides indications of a continuous phase transition.
Moreover, we identify the elementary excitation of the dynamics, a parameter responsible for allowing the particles to diffuse. We also discuss the possibility of topological defects affecting the transport of particles along the dynamics. Our findings conclude that the transition from limited to unlimited energy growth is of second order, hence a continuous phase transition.

%Remark: The normal component of the momentum is not conserved

\section{The model}
\begin{figure}[!hbt]
\begin{center}
\includegraphics[clip,angle=0,width=1\hsize]{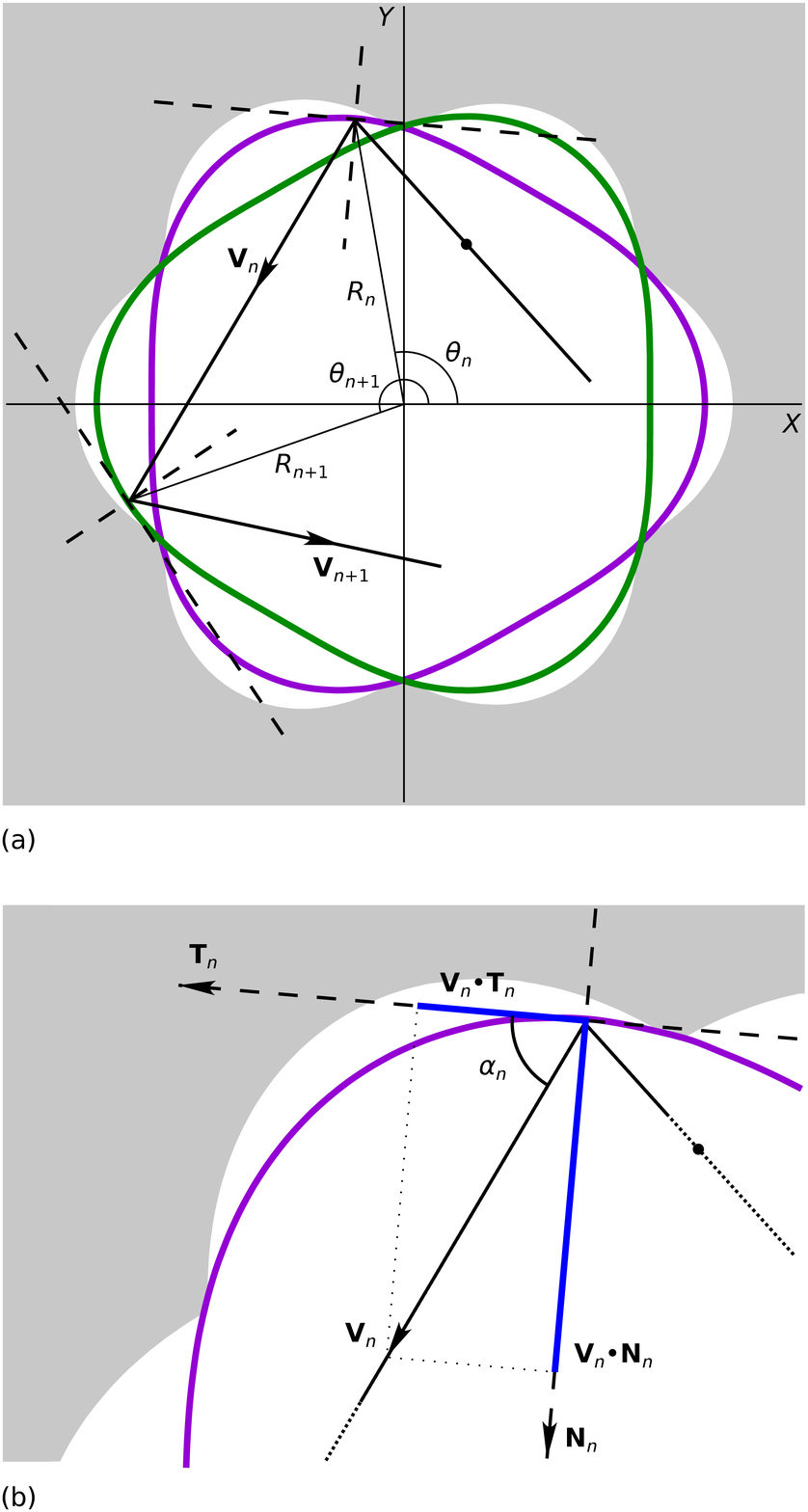}%{a.pdf}
\end{center}
\caption{\footnotesize{
(a) Piece of a particle's trajectory in the billiard under investigation. 
(b) Amplification of a portion on the top left of (a). 
}}
\label{fronteira}
\end{figure}

The main goal of this paper is to characterize a transition from
limited to unlimited energy gain for an ensemble of non-interacting particles confined within an oval billiard whose boundary oscillates periodically in time. The equation of the border is given by
\begin{equation}
R(\theta,\varepsilon, \eta, t, p) = 1 + \varepsilon\left[ 1 +\eta\cos(t) \right]\cos(p\theta),
\label{eqBoundary}
\end{equation}
where $p$ is an integer, and $\varepsilon$ is a parameter responsible for the geometrical deformation. For $\varepsilon = 0$, the billiard is a circle. The control parameter $\eta$ determines the amplitude of the time perturbation of the boundary, and the case $\eta = 0$ recovers the static billiard. The dynamics of a particle is specified in terms of its velocity $V_n$, the angular position $\theta_n$, the angle $\alpha_n$ that its trajectory forms with a tangent line at the position of the collision, and the time of the collision $t_n$ as
\begin{eqnarray}
X(t) =& X(\theta_n, t_n) + V_n\cos(\alpha_n+\phi_n)(t-t_n),\\
Y(t) =& Y(\theta_n, t_n) + V_n\sin(\alpha_n+\phi_n)(t-t_n),
\end{eqnarray}
%\begin{equation}
%X(t) = X(\theta_n, t_n) + V_n\cos(\alpha_n+\phi_n)(t-t_n),
%\end{equation}
%\begin{equation}
%Y(t) = Y(\theta_n, t_n) + V_n\sin(\alpha_n+\phi_n)(t-t_n),
%\end{equation}
where the time $t \geq t_n$ with $X(\theta_n, t_n) = R(\theta_n, t_n)\cos(\theta_n)$ and $Y(\theta_n, t_n) = R(\theta_n, t_n)\sin(\theta_n)$.
%The phase space for the time-dependent oval billiard is four-dimensional, having $\theta$, $t$, $V$, and $\alpha$ as independent variables. 
Once the angle $\theta$ is known, the angle $\phi$ is obtained and corresponds to the angle formed between the tangent and horizontal lines at the position $X(\theta)$, $Y(\theta)$, which can be written as $\phi = \arctan\left( Y'(\theta,t)/X'(\theta,t) \right) $ where $Y'$ and $X'$ are derivatives with respect to $\theta$. 
%Fig. \ref{fronteira} shows a typical illustration of the billiard and the angles used to describe the dynamics of the model.
Figure \ref{fronteira} illustrates the billiard under investigation for $\varepsilon=10^{-2}$, $\eta=20$ and $p=3$. The green and violet curves represent the boundary of the billiard at two instants. Figure \ref{fronteira}(a) depicts a portion of the particle's trajectory. 
At instant $t_n$, the particle collides against the boundary and acquires velocity ${\bf V}_n$. The point of the $n$th collision is characterized by the polar coordinates $(R_n, \theta_n)$. After a time interval traveling in a straight line, the particle reaches the boundary at instant $t_{n+1}$ and, immediately after the collision, its velocity is ${\bf V}_{n+1}$.
The gray region represents the portion of space outside the time-dependent boundary. 
Figure \ref{fronteira}(b) corresponds to an amplification of the top left portion of the billiard illustrated in Fig. \ref{fronteira}(a), where the particle undergoes the $n$th collision. In this figure, we use dashed lines to represent the tangent and normal directions relative to the boundary at the instant of the $n$th collision. 
This figure also presents the variable $\alpha_n$, which is the smaller angle between the vector velocity and the line tangent to the boundary. We also include in Figure \ref{fronteira}(b) the components of ${\bf V}_{n}$ relative to the tangent and normal directions with blue color. 
 Considering that the particle travels with constant speed between collisions, its position within the billiard is given in polar coordinates as $R_p(t)=\sqrt{X^2(t)+Y^2(t)}$. The angular position $\theta$ at the instant of impact is found numerically through 
the equation $R_p(\theta_{n+1},t_{n+1}) = R(\theta_{n+1}, t_{n+1})$. 

The particle's velocity has two components, a tangential and a normal. If the collisions of the particle with the boundary are inelastic partial loss of energy is observed, affecting, according to the reflection law, only the normal component of the velocity. At the instant of collision, the reflection law is written as
\begin{eqnarray}
\textbf{V}'_{n+1} \cdot \textbf{T}_{n+1} =& \textbf{V}'_n \cdot \textbf{T}_{n+1}, \\
\textbf{V}'_{n+1} \cdot \textbf{N}_{n+1} =& -\gamma\textbf{V}'_n \cdot \textbf{N}_{n+1},
\end{eqnarray}
%\begin{equation}
%\overrightarrow{V}'_{n+1} \cdot \overrightarrow{T}_{n+1} = \overrightarrow{V}'_n \cdot \overrightarrow{T}_{n+1},
%\end{equation}
%\begin{equation}
%\overrightarrow{V}'_{n+1} \cdot \overrightarrow{N}_{n+1} = -\gamma\overrightarrow{V}'_n \cdot \overrightarrow{N}_{n+1},
%\end{equation}
where $\gamma \in [0,1]$ is the restitution coefficient. If $\gamma=1$, the collisions are completely elastic, allowing the system 
to exhibit Fermi acceleration as the LRA conjecture states, whereas $\gamma<1$ leads to a fractional energy loss of the particles. 
The term $\bf{V}'$ corresponds to the velocity of the particle in the non-inertial reference frame of the boundary; the normal and tangent unit vectors are given, respectively, by
\begin{eqnarray}
\textbf{N}_{n+1} =& -\sin(\phi_{n+1})\textbf{i}+\cos(\phi_{n+1})\textbf{j}, \\
\textbf{T}_{n+1} =& \cos(\phi_{n+1})\textbf{i} + \sin(\phi_{n+1})\textbf{j}.
\end{eqnarray}
%\begin{equation}
%\overrightarrow{N}_{n+1} = -\sin(\phi_{n+1})\hat{i}+\cos(\phi_{n+1})\hat{j},
%\end{equation}
%\begin{equation}
%\overrightarrow{T}_{n+1} = \cos(\phi_{n+1})\hat{i} + \sin(\phi{n+1})\hat{j}.
%\end{equation}
After the impact, $n+1$, the tangential and normal velocity components are given by
\begin{eqnarray}
\textbf{V}_{n+1} \cdot \textbf{T}_{n+1} =& \textbf{V}_n \cdot \textbf{T}_{n+1}, \\
\textbf{V}_{n+1} \cdot \textbf{N}_{n+1} =& -\gamma\textbf{V}_n \cdot \textbf{N}_{n+1} \nonumber +\\
+&(1+\gamma)\textbf{V}_b (t_{n+1}+Z(n)) \cdot \textbf{N}_{n+1}, \label{sb}
\end{eqnarray}
%\begin{equation}
%\overrightarrow{V}_{n+1} \cdot \overrightarrow{T}_{n+1} = \overrightarrow{V}_n \cdot \overrightarrow{T}_{n+1},
%\end{equation}
%\begin{equation}
%\overrightarrow{V}_{n+1} \cdot \overrightarrow{N}_{n+1} = -\gamma\overrightarrow{V}_n \cdot \overrightarrow{N}_{n+1} +
%(1+\gamma)\overrightarrow{V}_b(t_{n+1}+Z(n)) \cdot \overrightarrow{N}_{n+1},
%\end{equation}
where $\textbf{V}_b (t_{n+1}+ Z(n))$ corresponds to the boundary velocity given by
\begin{equation}
\textbf{V}_b = \frac{dR}{dt}\Big|_{t_{n+1}+Z(n)} \left[ \cos(\theta_{n+1})\textbf{i} + \sin(\theta_{n+1}) \textbf{j} \right].
\end{equation}

The term $Z(n)$ is a random number between $0$ and $2\pi$ introduced to create a stochastic behavior in the movement of the oscillating boundary. The introduction guarantees that the $\alpha \times \theta$ plane is covered uniformly, i.e., the chaotic orbit can diffuse in all parts of the phase space. The presence of stochasticity does not change the velocity distribution for the dissipative case. Our results are, for this case, also applicable to the non-stochastic billiard. As a side remark, we mention that the stochastic oval billiard can be considered as a model for heat transfer of a gas \cite{Termal}, as a potential application.

Finally, the velocity of the particle and the angle $\alpha$ at the collision $(n+1)$ are given, respectively, by
\begin{equation}
V_{n+1}=\sqrt{(\textbf{V}_{n+1} \cdot \textbf{T}_{n+1})^2 + (\textbf{V}_{n+1} \cdot
	\textbf{N}_{n+1})^2},
\end{equation}
\begin{equation}
\alpha_{n+1}=\arctan\left[\frac{\textbf{V}_{n+1}\cdot\textbf{N}_{n+1}}{\textbf{V}_{n+1}\cdot\textbf{T}_{n+1}} \right].
\end{equation}
With the above equations, the description of the system is complete.

The dissipative oval billiard has a transition from limited for $\gamma\neq 1$ to unlimited energy growth for $\gamma=1$ \cite{LRA2,dissp}. It has been shown that for restitution parameter values close to but different from one, the dissipation is sufficient to prevent the particle from showing FA. When $\gamma$ is increased to reach the critical value, the system displays scale invariance, indicating the occurrence of a phase transition. Since our goal is to characterize a transition from limited to unlimited energy growth, we focus on the discussion of four principal items: 1) Identification of a broken symmetry; 2) definition of an order parameter; 3) discussion of the elementary excitations; and 4) discussion of the topological defects impacting the diffusion of the energy.

\section{The phase transition}

Three parameters control the system: $\eta$, associated with the movement of the boundary, $\varepsilon$, related to the amplitude of the circle deformation, and $\gamma$, denoting the restitution parameter. For $\eta=0$, the billiard is static, and the system is non-integrable. Depending on $\varepsilon$ and the initial conditions, chaotic components are observed in the phase space, and corresponding FA is expected to occur in the driven billiard. This holds for values of the parameter $\varepsilon \neq 0$ because the system turns into the circular billiard if $\varepsilon = 0$, whose static version is integrable. Therefore, for $\eta\varepsilon\neq 0$, the $\theta \times \alpha$ plane displays a chaotic sea, and chaotic diffusion is observed in the dynamics.

\begin{figure}[!hbt]
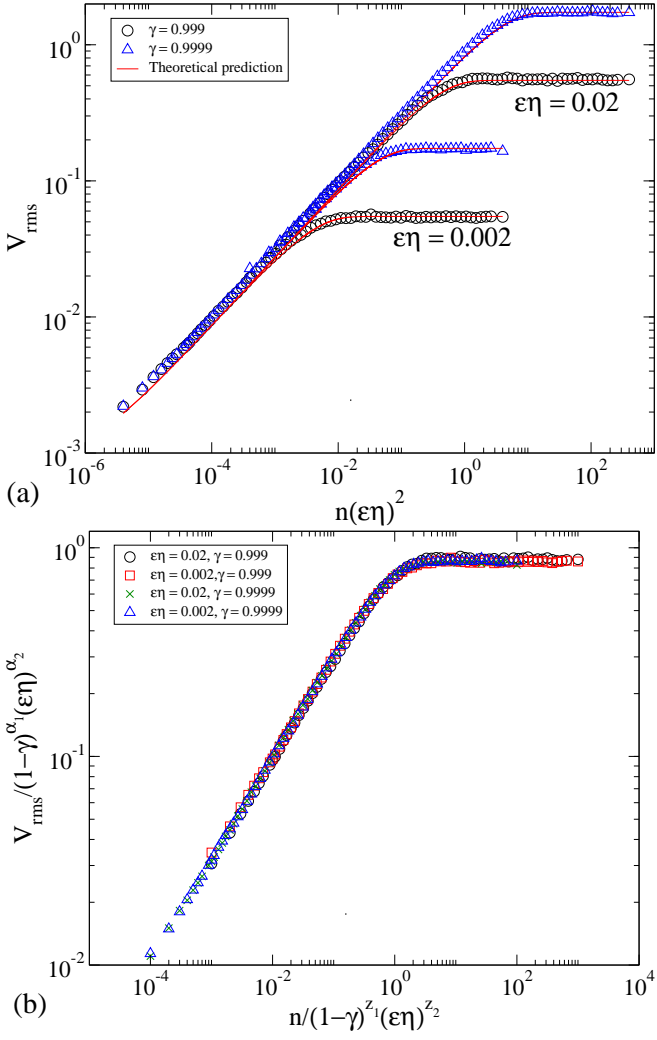

\begin{center}
\includegraphics[clip,angle=0,width=1\hsize]{Fig2a.eps}
\includegraphics[clip,angle=0,width=1\hsize]{Fig2b.eps}
\end{center}
\caption{\footnotesize{a) $V_{rms}$ vs. $n(\eta\varepsilon^2)$ for different values of $\gamma$ close to the value one and two
 combinations of $\eta\varepsilon$. The circles represent the numerical results, while the solid lines were obtained analytically. b) Overlap of the $V_{rms}$ curves after the following scaling transformatios: (i) $V\rightarrow V/((1-\gamma)^{\alpha_1}(\eta\varepsilon)^{\alpha_2})$; (ii) $n\rightarrow n/((1-\gamma)^{z_1}(\eta\varepsilon)^{z_2})$.}}
\label{curvasAna}
\end{figure}

The natural observable along the chaotic dynamics to prove the existence of the diffusion is the square root of the averaged
squared velocity, given by
\begin{equation}
V_{rms}=\sqrt{\frac{1}{M}\sum_{i=1}^{M}\frac{1}{n}\sum_{j=1}^{n}V^2_{i,j}},
\end{equation}
where $M$ corresponds to the number of initial conditions whereas $n$ is the number of collisions of the particle with the
boundary. As discussed in reference \cite{dissp}, the behavior of $V_{rms}$ is described as follows and as shown in Figure \ref{curvasAna}.
For relatively low initial velocity, $V_0\approx 0$, the curves of $V_{rms}$ grow as $V_{rms}\propto [n(\eta\varepsilon)^2]^\beta$ with $\beta\cong 1/2$ yielding the diffusion of particles in the velocity space to be equivalent to a normal diffusion. For large enough $n$, the curves bend towards a regime of saturation given by a constant plateau \cite{Termal}, marking a limitation for diffusion $V_{sat}\propto (1-\gamma)^{\alpha_1}(\eta\varepsilon)^{\alpha_2}$ with $\alpha_1=(-1/2)$, $\alpha_2=1$. The changeover from growth to the saturation is written as $n_x\propto (1-\gamma)^{z_1}(\eta\varepsilon)^{z_2}$, $z_1=-1$ and $z_2=0$. Using proper scaling, all curves shown in Fig. \ref{curvasAna}(a) fall onto each other in a single and universal plot confirming a scale invariance for the chaotic diffusion, as shown in Fig. \ref{curvasAna}(b).

The parameter controlling the criticality of the system is $\gamma$. For $\gamma=1$, FA can be observed, whereas $\gamma<1$ leads to the suppression of the energy gained by the particle. As a consequence, FA is not observed anymore. The order parameter goes continuously to zero for a second-order phase transition while its susceptibility diverges in the same limit. In previous works \cite{Termal,dissp2}, a set of critical exponents was provided using a phenomenological approach and considering a set of
scaling hypotheses allied with a homogeneous function. The plateau marking the saturation regime for the root mean square of the squared average velocity is given by $V_{rms,sat} \propto (1-\gamma)^{-1/2}$, therefore diverging in the limit of $\gamma\rightarrow 1$. However, an observable defined as $\sigma=1/V_{rms,sat}\propto \sqrt{1-\gamma}$ is eligible as an order parameter. It goes continuously to zero in the limit $\gamma\rightarrow 1$. Its susceptibility, defined as $\chi={{\partial\sigma}\over{\partial\gamma}}$ diverges in the same limit.

Let us now discuss the results for the average squared velocity. For the stochastic model, in which we consider the random number $Z$ in the argument of the velocity of the wall, the probability distribution for the velocity in the two-dimensional phase space $\alpha$ versus $\theta$ is uniform. It allows us to assume the statistical independence between the velocity and the dynamical variables $\theta$ and $\alpha$. In this case, taking the average of the velocity for the ranges $\alpha\in [0,\pi]$, $\theta\in [0,\pi]$ and $t\in [0,2\pi ]$ leads to 
\begin{equation}
\bar{V}^2_{n+1}=\frac{\bar{V}^2_n}{2}+\frac{\gamma^2\bar{V}^2_n}{2}
+\frac{(1+\gamma)^2\eta^2\varepsilon^2}{8}.
\end{equation}
Although the random number $Z$ becomes irrelevant when talking about $|\bf{V}_{n+1}|$ for the range $t\in[0,2\pi]$, its addition to the argument of the velocity of the wall is necessary. The phase space $\alpha$ versus $\theta$ is uniform, which is a necessary condition to obtain $\bar{V}^{2}_{n+1}$. The average squared velocity can be obtained assuming that 
$\bar{V}^{2}_{n+1}-\bar{V}^2_n=\frac{\bar{V}^2_{n+1}-\bar{V}^2_n}{(n+1)-n}\cong \frac{d\bar{V}^2}{dn}=\frac{\bar{V}^2(\gamma^2-1)}{2}+\frac{(1+\gamma)^2\eta^2\varepsilon^2}{8}$, where the differential equation has the following solution
\begin{equation}
\bar{V}^2(n)=\bar{V_0}^2e^{\frac{(\gamma^2-1)}{2}n}+\frac{(1+\gamma)}	{4(1-\gamma)}\eta^2\varepsilon^2\left[1-e^{\frac{(\gamma^2-1)}{2}n}\right].
\end{equation}
To compare the analytical prediction with the simulations, we must
average $\bar{V}^2$ over the orbit, which leads to
\begin{eqnarray}
\langle\bar{V}^2(n)\rangle=\frac{1}{n+1}\sum^n_{i=0}\bar{V}^2(i)=
\frac{(1+\gamma)}{4(1-\gamma)}\eta^2\varepsilon^2+  \nonumber\\
+\frac{1}{n+1}\left(\bar{V_0}^2-\frac{(1+\gamma)}{4(4-\gamma)
\eta^2\varepsilon^2}\right)\left[
\frac{1-e^{\frac{(\gamma^2-1)(n+1)}{2}}}{1-e^{\frac{(\gamma^2-1)}{2}}}\right].
\end{eqnarray}

\begin{figure}[!hbt]
\begin{center}
\includegraphics[clip,angle=0,width=1\hsize]{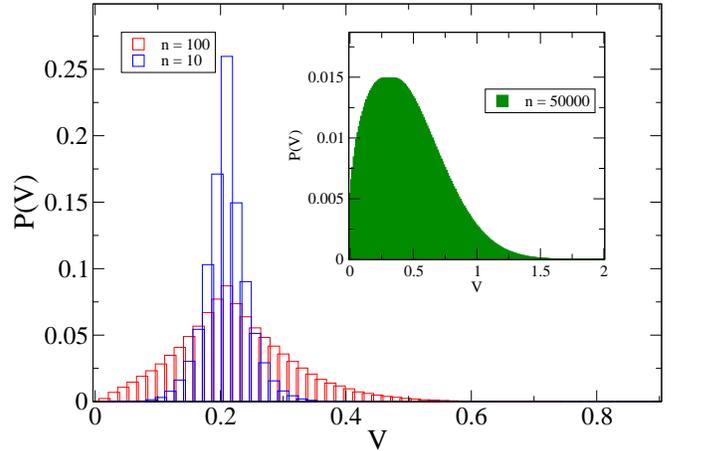}
\end{center}
\caption{\footnotesize{Histogram of the normalized probability distribution for the velocity for an ensemble of $10^5$ particles in the stochastic and dissipative oval billiard. The blue (dark gray) bars are obtained after ten collisions, while red bars (light gray) correspond to 100 collisions. The inset plot is obtained after 50000 collisions. The initial velocity is $V_0 = 0.2$ and the control parameters are $\eta=0.02$ and $\gamma=0.999$ with $p=2$.}}
\label{histo}
\end{figure}

Figure \ref{histo} shows the normalized probability distribution $P(V)$ for the three dynamical regimes: short, intermediate, and large time ranges.
The parameters used are $\eta=0.02$ and $\gamma=0.999$.
For larger times, a presumed Gaussian distribution flattens until the left-hand side of the curve reaches the lower velocity limit. The lower limit is given by the lowest velocity of the moving boundary. This change in probability behavior to a non-Gaussian distribution is similar to the one already observed in the Fermi-Ulam model \cite{leonel6}.

Figure \ref{curvasAna}(a) shows the dependence of $V_{rms}$ vs. $n$ for different values of $\varepsilon \eta$, as labeled in the figure. The range of parameters we are interested in is $\gamma$ close to one, specifically $\gamma\in[0.999, 0.99995]$ and small $\varepsilon\eta\in[0.002, 0.02]$. The last interval was chosen to facilitate numerical simulations, but other positive values would give similar results. The symbols correspond to the numerical simulation obtained from the mapping iteration considering an ensemble of $M=5\times 10^3$ different initial conditions starting with the same initial velocity $V_0=0.001$. The solid lines denote the analytical results. As we can see, the agreement with the numerical curves is excellent.

Considering now the limit of $n\rightarrow\infty$, the saturation for the average velocity gives 
 \begin{equation}
 V_{rms,sat}=\sqrt{\frac{1+\gamma}{1-\gamma}}\frac{\eta\varepsilon}{2}.
 \label{vsat}
 \end{equation}
As discussed earlier, the order parameter $\sigma$ is then given by
\begin{equation}
\sigma={{1}\over{V_{rms,sat}}}=\sqrt{{{1-\gamma}\over{1+\gamma}}}{{2}\over{\eta\varepsilon}}.
\end{equation}
We notice $\sigma$ depends on two sets of control parameters: (i) $\gamma$, which brings the criticality for the dynamics and makes $\sigma\rightarrow 0$ when $\gamma\rightarrow 1$, and; (ii) $\eta\varepsilon$, which by the range of control parameters considered does not bring criticality to the dynamics.

\begin{figure}[!hbt]
\begin{center}
\includegraphics[clip,angle=0,width=1\hsize]{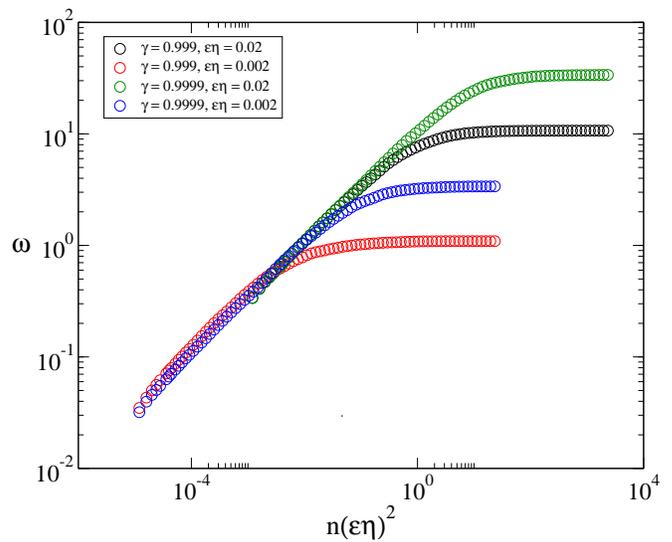}
\end{center}
\caption{\footnotesize{Standard deviation of $V_{rms}$ vs. $n$ for different values of the control parameters $\gamma$ and $\varepsilon\eta$. The horizontal axis is chosen to show the growth exponent is the same for all curves. Moreover, the standard deviation has the same critical exponents as the $V_{rms}$ curves.}}
\label{rugosidade}
\end{figure}

Let us now determine the expression of the susceptibility $\chi$, which gives information on how the order parameter responds to a variation of the control parameter $\gamma$ responsible for the criticality. It is defined as
\begin{equation}
\chi={{\partial\sigma}\over{\partial\gamma}}=-{{2}\over{\eta\varepsilon(1+\gamma)^2}}\sqrt{{{1+\gamma}\over{1-\gamma}}},
\end{equation}
and it diverges in the limit $\gamma\rightarrow 1$. The behavior of the two observables $\sigma$ and $\chi$ for the limit $\gamma\rightarrow 1$ provides evidence that this is a continuous phase transition. Moreover, the fluctuations can be analyzed through the standard deviation of the velocity over $M$ initial conditions
\begin{equation}
\omega(\eta\varepsilon,n)=\frac{1}{M}\sum_{j=1}^M\sqrt{\overline{V^2}_j(\eta\varepsilon,n,V_0)-\overline{V}^2_j(\eta\varepsilon,n,V_0)}. 
\end{equation}
Figure \ref{rugosidade} shows the behavior of the standard deviation $\omega$ for the dissipative case. Similar to $V_{rms}$, $\omega$ has a growth regime for low values of $n$ and reaches a plateau after many collisions. Therefore, the fluctuations do not grow unbounded as they would for the non-dissipative case. There is a rather limited range of fluctuations, which varies depending on the parameters $\eta$ and $\varepsilon$. It indicates that the long-range ``interactions'' during the phase transition are limited, which is expected as the dissipation
leads to a phase space contraction, interrupting the FA. 
  
Let us discuss the symmetry breaking taking place during the phase transition. In the limit of high velocities $V_n \rightarrow\infty$ of the particles such that $\bf{V_b}$ is negligible, the normal component of the velocity given by equation (\ref{sb}) would not change its absolute value upon each collision for the $\gamma=1$ dynamics. This symmetry of equation (\ref{sb}) is broken for $\gamma<1$ in the
sense that the absolute value of the velocity decreases after each collision and stays finite.
The change in the velocity behavior can be easily traced for the general case by comparing the system's dynamics in two situations: when the collisions are elastic vs. the collisions being inelastic.
\begin{figure}[!hbt]
\begin{center}
\includegraphics[clip,angle=0,width=1\hsize]{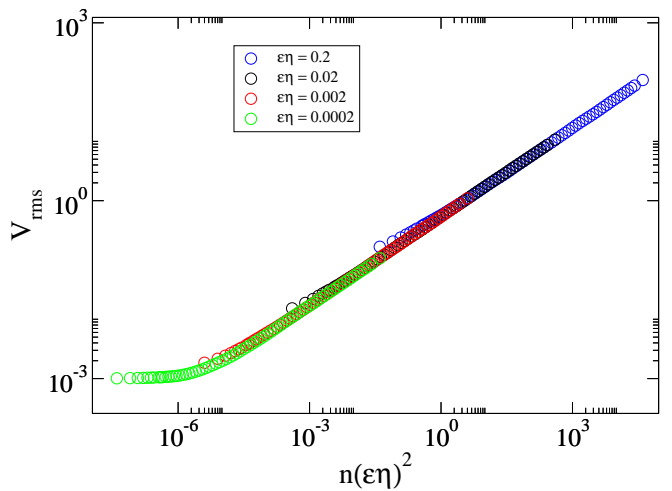}
\end{center}
\caption{\footnotesize{$V_{rms}$ vs. $n$ for the non-dissipative oval billiard, i.e, $\gamma=1$, for 4 different combinations of $\eta\epsilon$.}}
\label{nodiss}
\end{figure}

Figure \ref{nodiss} illustrates the behavior of $V_{rms}$ as a function of $n$ for different values of $\varepsilon\eta$ considering elastic collisions, i.e., $\gamma=1$. In this scenario, the system presents FA where, after a short number of collisions, the velocity increases monotonically with an exponent $\beta\approx 1/2$. However, for $\gamma<1$, the dissipation implies an area contraction of the accessible phase space leading to the creation of attractors. Given they are far away from infinity, unlimited diffusion is prevented. Hence, for those values of $\gamma$, the system shows a saturation regime in this phase, as shown in Fig. \ref{curvasAna}, demonstrating that the unlimited energy growth has been suppressed. Moreover, this discussion about the attractors takes us to the next point.
%Therefore, a symmetry break occurs near $\gamma=1$ when the collisions change from elastic to inelastic.

The next step regards a discussion of the existence of topological defects. In the conservative case, topological defects might be linked to periodic structures such as KAM islands \cite{edl}, where local trapping produced by stickiness happens. In the dissipative case, however, KAM islands are not present. The elliptical fixed points (or equivalently periodic orbits) centering the islands generally turn into sinks (attractive fixed points), attracting nearby initial conditions and trapping them. We have performed extensive simulations seeking the convergence of the dynamics in such regions, but we did not observe it. Therefore, we did not detect any attractive fixed points for the range of control parameters considered close to the transition point. 

Let us briefly discuss the elementary excitation responsible for the diffusion of the particles. Each particle of the ensemble's velocity changes after a collision with the boundary. In this context, the elementary excitation or the effect analogous to an elementary excitation occurs due to the product $\varepsilon\eta$, which is the maximal velocity of the billiard
boundary. Therefore $\varepsilon\eta$ defines the elementary unit of the underlining random walk ``motion'' of the trajectories in the phase space. Indeed, if we take the limit of short times and small initial velocities, the diffusion of the velocity is given by $V\approx \frac{\eta\varepsilon}{2}\sqrt{n}$. This shows that for a small $n$, the particle diffuses in the phase space analogously to a random walk with step size $\frac{\eta\varepsilon}{2}$. 

\section{Summary}

We have analyzed Fermi acceleration in a driven and stochastified oval billiard with and without
dissipation emphasizing its unlimited versus limited energy growth.
Our focus was hereby on the analogy to a phase transition taking place from the non-dissipative
to the dissipative dynamics. The suppression of infinite diffusion in momentum space
happens due to inelastic collisions leading the dynamics for a sufficiently long time
to approach a saturation regime. We identified an order parameter that goes continuously
to zero at the transition point and whose corresponding susceptibility diverges in the
same limit. The elementary excitation is given by the change of the velocity in an elementary
collision and we suggest it depends on the product of the involved parameters of the driven oval billiard. For the range of control parameters considered, topological defects were not detected.
Symmetry breaking occurs due to the non-accessibility of the high-velocity regime in the
presence of dissipation. Altogether this makes us conclude that there is a second-order phase transition
occurring when passing from unlimited to limited energy growth due to dissipation.

\section*{Acknowledgments}

 F.A.O.S acknowledges CAPES (No. 88887.670331/2022-00) for financial support. A.K.P.F. acknowledges FAPESP (2020/07219-1) for financial support. E.D.L. acknowledges support from Brazilian agencies CNPq (No. 301318/2019-0) and FAPESP (No. 2019/14038-6 and 2021/09519-5).


\begin{thebibliography}{99}


\bibitem{billiardWorks} N. S. Krylov, Works on the Foundations of Statistical Physics, Princeton University Press, Princeton, NJ, USA, 1979.

\bibitem{Chernov}N. Chernov, R. Markarian, Chaotic Billiards, vol. 127, American Mathematical Society, 2006

\bibitem{LRA1}A. Loskutov, A. B. Ryabov, and L. G. Akinshin, J. Exp. Theor. Phys. 89, 966 (1999).

\bibitem{LRA2}A. Loskutov, A. B. Ryabov, and L. G. Akinshin, J. Phys. A 33, 7973 (2000).

\bibitem{lorentzGas1}A. K. Karlis, P. K. Papachristou, F. K. Diakonos, V. Constantoudis, and P. Schmelcher, Phys. Rev. E 76, 016214 (2007).

\bibitem{lorentzGas2}D. F. M. Oliveira and E. D. Leonel, Chaos 22, 026123 (2012).

\bibitem{stadium}C. P. Dettmann and O. Georgiou, Phys. Rev. E 83, 036212 (2011).

\bibitem{ovalb}E. D. Leonel, D. F. M. Oliveira, and A. Loskutov, Chaos 19,
033142 (2009).

\bibitem{robusto1} F. Lenz, F. K. Diakonos, P. Schmelcher, Phys. Rev. Lett. 100 014103 (2008).

\bibitem{robusto2}F. Lenz, C. Petri, F. R. N. Koch, F.K. Diakonos, P. Schmelcher, New J. Phys. 11 083035 (2009).

\bibitem{robusto3}E. D. Leonel, L. A. Bunimovich, Phys. Rev. Lett. 104 224101 (2010).

\bibitem{Termal}E. D. Leonel, M. V. C. Galia and L. A. Barreiro, D. F. M. Oliveira, Phys. Rev. E {\bf 94}, 062211 (2016).

\bibitem{eins}M. L. Kavvas, R. S. Govindaraju, U. Lall, Chaos 25, 075201 (2015).

\bibitem{zwei}K. E. Chlouverakis, J. C. Scprott, Chaos 17, 023110 (2007).

\bibitem{drei}G. M. Zaslavsky, M. Edelman, B. A. Niayzov, Chaos 7, 150 (1997).

\bibitem{vier}Y. Pesin, M. Shlesinger, Y. Sinai, G. Zaslavsky, Chaos 7, 1 (1997).

\bibitem{funf}L. E. Reichl, A Modern Course in Statistical Physics, Wiley-VCH Verlag, Weinheim 2009.

\bibitem{sechs}A. L. Barabási, H. E. Stanley, Fractal Concepts in Surface Growth, Cambridge University Press, Cambridge 1985.

\bibitem{sieben}J. Cardy, Scaling and Renormalization in Statistical Physics, Cambridge University Press 2003.

\bibitem{arch}L. P. Kadanoff, Statistical Physics: Statics, Dynamics and Renormalization, World Scientific Publishing Co. Pte. Ltd., Singapore 2000.

\bibitem{neun}E. D. Leonel, P. V. E. McClintock, J. K. L. da Silva, Phys. Rev. Lett. 93, 014101 (2004).

\bibitem{zehn}D. G. Ladeira, J. K. L. da Silva, J. Phys. A 41, 365101 (2008).

\bibitem{dissp}Diego F. M. Oliveira, E. D. Leonel, Physica A 389, 1009 (2010).

\bibitem{dissp2}Edson D. Leonel, Carl P. Dettmann, Phys. Lett. A 376, 1669 (2012).

\bibitem{leonel6}D. F. M. Oliveira, Mario R. Silva, and E. D. Leonel, Physica A 436, 909 (2015).

\bibitem{edl} E. D. Leonel, M. Yoshida, and J. A. Oliveira, EPL 131, 20002 (2020).


\end{thebibliography}
\end{document}